\documentclass[10pt]{artikel3}
\usepackage{amsfonts,amsmath,amssymb}
\usepackage{float,braket}
\usepackage{pslatex}
\usepackage{subfigure}
\usepackage{graphicx,wrapfig}

\newcommand{\MSbar}{\overline{\mbox{MS}}}

\newcommand{\p}{\partial}

\newcommand{\omu}{\overline{\mu}}

\newcommand{\oB}{\overline{B}}
\newcommand{\occ}{\overline{c}}
\newcommand{\oG}{\overline{G}}

\renewcommand{\d}{\ensuremath{\mathrm{d}}}
\newcommand{\e}{\ensuremath{\mathrm{e}}}
\usepackage[small,bf]{caption}
\newcommand{\sect}[1]{ \section{#1} \setcounter{equation}{0} }

\begin{document}
\title{{\bf A potential setup for perturbative confinement}}
\author{David Dudal\thanks{david.dudal@ugent.be}\\\\
\small \textnormal{Center for Theoretical Physics, Massachusetts Institute of Technology,} \\\small \textnormal{77 Massachusetts Avenue, Cambridge, MA 02139, USA}\\\\
\small \textnormal{Ghent University, Department of Mathematical
Physics and Astronomy,} \\ \small \textnormal{Krijgslaan 281-S9,
9000 Gent, Belgium}\normalsize}

\date{}
\maketitle
\begin{abstract}
\noindent A few years ago, 't Hooft suggested a way to discuss
confinement in a perturbative fashion. The original idea was put
forward in the Coulomb gauge at tree level. In recent years, the
concept of a nonperturbative short distance linear potential also
attracted phenomenological attention. Motivated by these
observations, we discuss how a perturbative framework, leading to a
linear piece in the potential, can be developed in a manifestly
gauge and Lorentz invariant manner, which moreover enjoys the
property of being renormalizable to all orders. We provide an
effective action framework to discuss the dynamical realization of
the proposed scenario in Yang-Mills gauge theory.
\end{abstract}
\vspace{-11cm} \hfill MIT-CTP--4043 \vspace{10cm}

\sect{Motivation} In
\cite{'tHooft:2002wz,'tHooft:2003ih,'tHooft:2007zz}, 't Hooft
launched the idea that confinement can be looked upon as a natural
renormalization phenomenon in the infrared region of a Yang-Mills
gauge theory. He employed the Coulomb gauge, $\p_i A_i=0$, in which
case the kinetic (quadratic) part of the gauge field action becomes
\begin{equation}\label{1}
    S_{YM}=-\frac{1}{4}\int \d^4x F_{\mu\nu}^2\to \int
    \d^4x\left(-\frac{1}{2}(\p_i A_j)^2 +\frac{1}{2}(\p_0A_j)^2+\frac{1}{2}(\p_j
    A_0)^2\right)\,.
\end{equation}
The usual (classical) Coulomb potential is recovered as the solution
of the equation of motion for $A_0$ in the presence of static
charges with strength $\alpha_s$ (= source terms) separated from
each other by a vector $\mathbf{r}$,
\begin{equation}\label{2}
    V_{Q\overline{Q}}(\mathbf{r})=-\frac{\alpha_s}{r}\,.
\end{equation}
He then proposed that some (unspecified) infrared quantum effects
will alter the kinetic part into
\begin{equation}\label{3}
    \int
    \d^4x\left(-\frac{1}{2}(\p_i A_j)^2 +\frac{1}{2}(\p_0A_j)^2+\frac{1}{2}(\p_j
    A_0)^2\right)+\int \d^4x\left(-\frac{1}{2} \p_j A_0\frac{2\sigma/\alpha_s}{-\p_j^2+2\sigma/\alpha_s}\p_j
    A_0\right)\,.
\end{equation}
As a consequence, the Coulomb potential in momentum space gets
modified into
\begin{equation}\label{4}
    V_{Q\overline{Q}}(\mathbf{p})=-\frac{4\pi\alpha_s}{\mathbf{p}^2}-\frac{8\pi\sigma}{\mathbf{p}^4}\,,
\end{equation}
which corresponds to
\begin{equation}\label{5}
    V_{Q\overline{Q}}(\mathbf{r})=-\frac{\alpha_s}{r}+\sigma r\,,
\end{equation}
which is nothing else than a confining potential of the Cornell type
\cite{Eichten:1978tg}. We made use of the well-known identity
\mbox{$\p_i^2 \frac{1}{r}=-4\pi\delta(\mathbf{r})$}, which also
allows one to define a regularized version of the Fourier transform
of $\frac{1}{\mathbf{p}^4}$, since \mbox{$\p_i^2(r)=\frac{2}{r}$}.
Indeed, calling $f(\mathbf{p})$ the Fourier transform of $r$, we can
write
\begin{equation}\label{extraj1}
    \p_i^2 \p_i^2 \int\frac{\d^3\mathbf{p}}{(2\pi)^3}f(\mathbf{p})
    e^{i\mathbf{p}\cdot\mathbf{r}}=-8\pi \int\frac{\d^3\mathbf{p}}{(2\pi)^3}
    e^{i\mathbf{p}\cdot\mathbf{r}}\,,
\end{equation}
which leads to
\begin{equation}\label{extraj2}
    f(\mathbf{p})=-\frac{8\pi}{\mathbf{p}^4}\,.
\end{equation}
Of course, this is an appealing idea, at it might give a way to
handle confining theories in a relatively ``simple'' way, modulo the
fact that the origin of the parameter (= string tension) $\sigma$ is
still rather unclear. It was argued that the coefficient
$\frac{\sigma}{\alpha_s}$ has to be adjusted in such a way that
higher order corrections converge as fast as possible
\cite{'tHooft:2002wz,'tHooft:2003ih}.

In this work, we intent to set a modest step forward in this
program. First of all, we would like to avoid the use of a
non-Lorentz covariant gauge fixing as the Coulomb one, in fact, we
should rather avoid using any preferred gauge and produce a Lorentz
and gauge invariant version of the 't Hooft mechanism. Secondly, in
\cite{'tHooft:2002wz,'tHooft:2003ih} it was assumed that the
infrared effects would not reflect on the ultraviolet sector. Here,
we can even explicitly prove the ultraviolet renormalizability of
the procedure. We also point out shall how it would be possible to
dynamically realize this perturbative confinement scenario, starting
from the original Yang-Mills action.

Let us also refer to \cite{Chetyrkin:1998yr}, which gives a second
motivation for this work. In the phenomenological paper
\cite{Chetyrkin:1998yr}, the issue of physical $\frac{1}{q^2}$ power
corrections was discussed. Such $\frac{1}{q^2}$ corrections are in
principle forbidden to appear in the usual Operator Product
Expansion (OPE) applied to physical correlators, since there is no
local dimension 2 gauge invariant condensate to account for the
quadratic power correction. This wisdom was however challenged in
\cite{Chetyrkin:1998yr}, by including nonperturbative effects beyond
the OPE level. Next to the motivation based on ultraviolet
renormalons and/or approaches in which the Landau pole is removed
from the running coupling, which lead to $\frac{1}{q^2}$
uncertainties when studying the correlators, it was noticed that a
linear piece survives in the heavy quark potential up to short
distances. This means that a Cornell potential \eqref{5} could also
leave its footprints at distances smaller than might be expected. In
the meantime, the notion of a short distance linear potential has
also been discussed by means of the gauge/gravity duality approach
(AdS/QCD), see e.g. \cite{Andreev:2006ct,Zakharov:2007zzb}. Notice
hereby that the string tension at short distances does not have to
concur with the one at larger distances
\cite{Andreev:2006ct,Zakharov:2007zzb}.

\sect{Constructing the starting action and some of its properties}
We shall work in Euclidean space. We shall make a small detour
before arriving to our actual purpose of the note. We start from the
usual Yang-Mills action, and we couple the nonlocal gauge invariant
operator
\begin{equation}\label{detour1}
    \mathcal{O}(x)=F_{\mu\nu}^a(x)\left[\frac{1}{D_\rho^2}\right]^{ab}(x)F_{\mu\nu}^b(x)
\end{equation}
to it by means of a global ``source'' $J^2$, i.e. we consider
\begin{eqnarray}\label{detour1bis}
 S_{YM}+S_{\mathcal{O}}&=&\frac{1}{4}\int \d^4y F_{\mu \nu }^{a}F_{\mu \nu
 }^{a}-\frac{J^2}{4}\int \d^4x \mathcal{O}(x)\,.
\end{eqnarray}
This particular operator was first put to use in
\cite{Jackiw:1995nf,Jackiw:1997jga} in the context of a dynamical
mass generation for $3D$ gauge theories.

We introduced the formal notation $\frac{1}{D^2}$, which corresponds
to the (nonlocal) inverse operator of $D^2$, i.e.
\begin{equation}\label{extraj5}
    \frac{1}{D^2}(x)f(x)\equiv \int \d^4y \left[\frac{1}{D^2}\right](x-y)f(y)
\end{equation}
for a generic function $f(x)$, whereby
\begin{equation}\label{extraj6}
    D^2(x) \left[\frac{1}{D^2}\right](x-y)=\delta(x-y)\,.
\end{equation}
Imposing a gauge fixing by adding a gauge fixing term and
corresponding ghost part $S_{gf}$  to the action
\begin{equation}\label{detour2}
    S=S_{YM}+S_{\mathcal{O}}+S_{gf}\;,
\end{equation}
it was shown in \cite{Capri:2005dy,Capri:2006ne} that the partition
function,
\begin{equation}\label{detour3}
    \int[\d\Phi] e^{-S}\,,
\end{equation}
can be brought in a localized form by introducing a pair of complex
bosonic antisymmetric tensor fields $\left( B_{\mu \nu
}^{a},\overline{B}_{\mu \nu }^{a}\right)$ and of complex
anticommuting antisymmetric tensor fields $\left( \overline{G}_{\mu
\nu }^{a},G_{\mu \nu }^{a}\right)$, both belonging to the adjoint
representation, so that the nonlocal action $S$ gets replaced by its
equivalent local counterpart\footnote{Performing the Gaussian path
integration over $(B,\oB,G,\oG)$ leads back to \eqref{detour1bis}.}
\begin{eqnarray}
 S'&=&\int \d^4x\left[\frac{1}{4}F_{\mu \nu }^{a}F_{\mu \nu }^{a}+\frac{i}{4}J(B-\overline{B})_{\mu\nu}^aF_{\mu\nu}^a
  +\frac{1}{4}\left( \overline{B}_{\mu \nu
}^{a}D_{\sigma }^{ab}D_{\sigma }^{bc}B_{\mu \nu
}^{c}-\overline{G}_{\mu \nu }^{a}D_{\sigma }^{ab}D_{\sigma
}^{bc}G_{\mu \nu }^{c}\right)\right]\;, \label{detour4}
\end{eqnarray}
such that
\begin{equation}\label{detour3}
    \int[\d\Phi] e^{-S}=\int[\d\Phi] e^{-S'}\,.
\end{equation}
The shorthand notation $\Phi$ represents all the fields present in
$S$ or $S'$. The covariant derivative is given by
\begin{equation}\label{extraj3}
    D_{\mu}^{ab}=\delta^{ab}\p_\mu-gf^{abc}A_\mu^c\,.
\end{equation}
From now on, we can forget about the original starting point
\eqref{detour1bis}, and start our discussion from the local action
\eqref{detour4}, whereby $J$ can now also be considered to be a
local source $J(x)$, coupled to the operator
$(B-\overline{B})_{\mu\nu}^aF_{\mu\nu}^a$.

This is however not the end of the story. It was proven in
\cite{Capri:2005dy,Capri:2006ne} that $S'$ must be extended in order
to obtain a renormalizable action. More precisely, the complete
starting action is given by
\begin{eqnarray}
 \Sigma &=&\int \d^4x\left[\frac{1}{4}F_{\mu \nu }^{a}F_{\mu \nu }^{a}+\frac{iJ}{4}(B-\overline{B})_{\mu\nu}^aF_{\mu\nu}^a
  +\frac{1}{4}\left( \overline{B}_{\mu \nu
}^{a}D_{\sigma }^{ab}D_{\sigma }^{bc}B_{\mu \nu
}^{c}-\overline{G}_{\mu \nu }^{a}D_{\sigma }^{ab}D_{\sigma
}^{bc}G_{\mu \nu
}^{c}\right)\right.\nonumber\\
&&-\left.\frac{3}{8}%
J^{2}\lambda _{1}\left( \overline{B}_{\mu \nu }^{a}B_{\mu \nu
}^{a}-\overline{G}_{\mu \nu }^{a}G_{\mu \nu }^{a}\right)
+J^{2}\frac{\lambda _{2}}{32}\left( \overline{B}_{\mu \nu
}^{a}-B_{\mu \nu }^{a}\right) ^{2}\right.\nonumber\\&&\left.+
\frac{\lambda^{abcd}}{16}\left( \overline{B}_{\mu\nu}^{a}B_{\mu\nu}^{b}-\overline{G}_{\mu\nu}^{a}G_{\mu\nu}^{b}%
\right)\left( \overline{B}_{\rho\sigma}^{c}B_{\rho\sigma}^{d}-\overline{G}_{\rho\sigma}^{c}G_{\rho\sigma}^{d}%
\right) +\varsigma\,J^4\right]+S_{gf}\;, \label{completeactionb}
\end{eqnarray}
We shall clarify the significance of the vacuum term $\varsigma\,
J^4$, with $\varsigma$ a dimensionless parameter, after \eqref{ea1}.
$\lambda^{abcd}$ is an invariant rank 4 tensor coupling, subject to
the following symmetry constraints
\begin{eqnarray}
\lambda^{abcd}=\lambda^{cdab} \;, \lambda^{abcd}=\lambda^{bacd} \;,
\label{abcd}
\end{eqnarray}
which can be read off from the vertex that $\lambda^{abcd}$
multiplies \cite{Capri:2005dy,Capri:2006ne}.

In general, an invariant tensor $\lambda^{abcd}$ is defined by means
of \cite{vanRitbergen:1998pn}
\begin{equation}\label{extraj3}
    \lambda^{abcd}=\mathrm{Tr}(t^a t^b t^c t^d)\,,
\end{equation}
with $t^a$ the $SU(N)$ generators in a certain representation $r$.
\eqref{extraj3} is left invariant under the transformation
\begin{equation}\label{extraj4}
    t^a\to U^+ t^aU\,,\qquad U=\e^{i\omega^b t^b}\,,
\end{equation}
which leads for infinitesimal $\omega^a$ to the generalized Jacobi
identity \cite{vanRitbergen:1998pn}
\begin{equation}\label{jacobigen}
    f^{man}\lambda^{mbcd}+f^{mbn}\lambda^{amcd}+f^{mcn}\lambda^{abmd}+f^{mdn}\lambda^{abcm}=0\,.
\end{equation}
It are the radiative corrections which necessitate the introduction
of the extra terms $\propto \lambda_{1,2}J^2$, as well as the
quartic interaction $\propto\lambda^{abcd}$
\cite{Capri:2005dy,Capri:2006ne}. The quantities $\lambda_1$ and
$\lambda_2$ are two a priori independent scalar ``couplings''.

It can be easily checked that \eqref{completeactionb} is gauge
invariant, $\delta_\omega S=0$, w.r.t. to the infinitesimal gauge
variations
\begin{eqnarray}
\delta_\omega A_{\mu }^{a} =-D_{\mu }^{ab}\omega ^{b}\;,
\delta_\omega B_{\mu \nu }^{a} =gf^{abc}\omega ^{b}B_{\mu \nu
}^{c}\;, \delta_\omega \overline{B}_{\mu \nu }^{a} =gf^{abc}\omega
^{b}\overline{B}_{\mu \nu }^{c}\;, \delta_\omega G_{\mu \nu }^{a}
=gf^{abc}\omega ^{b}G_{\mu \nu }^{c}\;,  \delta_\omega
\overline{G}_{\mu \nu }^{a} =gf^{abc}\omega ^{b}\overline{G}_{\mu
\nu }^{c}\;. \label{gtm}\nonumber\\
\end{eqnarray}
Using a linear covariant gauge,
\begin{eqnarray}
  S_{gf}&=&\int \d^{4}x\;\left( \frac{\alpha }{2}b^{a}b^{a}+b^{a}%
\partial _{\mu }A_{\mu }^{a}+\overline{c}^{a}\partial _{\mu }D_{\mu
}^{ab}c^{b}\right)\;,\label{lcg}
\end{eqnarray}
it was shown in  \cite{Capri:2005dy,Capri:2006ne} that the action
$\Sigma$, \eqref{completeactionb}, is renormalizable to all orders
of perturbation theory, making use of the algebraic formalism and
BRST cohomological techniques \cite{Piguet:1995er}. Indeed, the
action \eqref{completeactionb} enjoys a nilpotent BRST symmetry,
generated by
\begin{eqnarray}\label{brst1}
sA_{\mu }^{a} &=&-D_{\mu }^{ab}c^{b}\;,  sc^{a}
=\frac{g}{2}f^{abc}c^{b}c^{c}\;,  sB_{\mu \nu }^{a}
=gf^{abc}c^{b}B_{\mu \nu }^{c}\;, s\overline{B}_{\mu \nu }^{a}
=gf^{abc}c^{b}\overline{B}_{\mu \nu }^{c}\;,\nonumber \\sG_{\mu \nu
}^{a} &=&gf^{abc}c^{b}G_{\mu \nu }^{c}\;, s\overline{G}_{\mu \nu
}^{a} =gf^{abc}c^{b}\overline{G}_{\mu \nu }^{c}\;, s\overline{c}^{a}
=b^{a}\;, sb^{a} =0\;,s^2=0, s\Sigma=0\,.
\end{eqnarray}
Later on, the renormalizability was also confirmed in the more
involved maximal Abelian gauge \cite{Capri:2007ph}.

If we put the source $J=0$, we expect to recover the usual
Yang-Mills theory we started from, see \eqref{detour1bis}. Though,
the action \eqref{completeactionb} with $J=0$,
\begin{eqnarray}
 S_{YM}' &=&\int \d^4x\left[\frac{1}{4}F_{\mu \nu }^{a}F_{\mu \nu }^{a}+\frac{1}{4}\left( \overline{B}_{\mu \nu
}^{a}D_{\sigma }^{ab}D_{\sigma }^{bc}B_{\mu \nu
}^{c}-\overline{G}_{\mu \nu }^{a}D_{\sigma }^{ab}D_{\sigma
}^{bc}G_{\mu \nu }^{c}\right)\right.\nonumber\\&&\left.+
\frac{\lambda^{abcd}}{16}\left( \overline{B}_{\mu\nu}^{a}B_{\mu\nu}^{b}-\overline{G}_{\mu\nu}^{a}G_{\mu\nu}^{b}%
\right)\left( \overline{B}_{\rho\sigma}^{c}B_{\rho\sigma}^{d}-\overline{G}_{\rho\sigma}^{c}G_{\rho\sigma}^{d}%
\right)\right]\;, \label{completeactionc}
\end{eqnarray}
seems to differ from the ordinary gluodynamics action $S_{YM}$. This
is however only apparent. Following
\cite{Capri:2006ne,Dudal:2007ch}, we introduce the nilpotent
``supersymmetry'' $\delta^{(2)}$,
\begin{eqnarray}\label{ss}
\delta^{(2)} B_{\mu\nu}^a &=& G_{\mu\nu}^a \;,\delta^{(2)}
G_{\mu\nu}^a =0 \;, \delta^{(2)} \overline{G}_{\mu\nu}^a =
\overline{B}_{\mu\nu}^a \;, \delta^{(2)} \overline{B}_{\mu\nu}^a =
0\;, \delta^{(2)}\delta^{(2)}=0\;, \delta^{(2)}
\left(S_{YM}'+S_{gf}\right)=0\;.
\end{eqnarray}
based on which it can be shown that the newly introduced tensor
fields $\{B_{\mu\nu}^a,{\overline B}_{\mu\nu}^a, G_{\mu\nu}^a,
{\overline G}_{\mu\nu}^a\}$ do not belong to the cohomology of
$\delta^ {(2)}$, as they constitute pairs of
$\delta^{(2)}$-doublets, and as such completely decouple from the
physical spectrum \cite{Piguet:1995er}. This means that $S_{YM}$ and
$S_{YM}'$ share the same physical degrees of freedom, being 2
transverse gluon polarizations, as can be proven using the BRST
cohomology \cite{Dudal:2007ch}.

In addition, the tensor coupling $\lambda^{abcd}$ cannot enter the
Yang-Mills correlators constructed from the original Yang-Mills
fields $A_\mu^a, b^a, \occ^a, c^a$ as it is
coupled to a $\delta^{(2)}$-exact term, $\propto \delta^{(2)}\left[\left( \overline{B}_{\mu \nu }^{a}B_{\mu \nu }^{b}-\overline{G}_{\mu \nu }^{a}G_{\mu \nu }^{b}%
\right)\left(\overline{G}_{\rho\sigma}^{c}B_{\rho\sigma}^d\right)\right]$,
hence $\lambda^{abcd}$ w.r.t. Yang-Mills correlators plays a role
akin to that of a gauge parameter w.r.t. gauge invariant
correlators.

The gauge invariant action $S_{YM}'$, \eqref{completeactionc}, is
thus perturbatively completely equivalent with the usual Yang-Mills
action: it is renormalizable to all orders of perturbation theory,
and the physical spectrum is the same. The advantage of $S_{YM}'$ is
that it allows to couple a gauge invariant local composite operator
to it, which is written down in \eqref{completeactionb}. This means
that we can probe Yang-Mills gauge theories with this particular
operator, and investigate the associated effective action, to find
out whether a gauge invariant condensate is dynamically favoured.

\sect{The effective action for the gauge invariant operator
$(B_{\mu\nu}^a-\oB_{\mu\nu}^a)F_{\mu\nu}^a$ } We consider the
functional $W(J)$, given by
\begin{equation}\label{ea1}
    e^{-W(J)}=\int [\d\Phi] e^{-S_{YM}'-\int \d^4x\left(\frac{iJ}{4}(B-\overline{B})_{\mu\nu}^aF_{\mu\nu}^a
  -\frac{3}{8}%
J^{2}\lambda _{1}\left( \overline{B}_{\mu \nu }^{a}B_{\mu \nu
}^{a}-\overline{G}_{\mu \nu }^{a}G_{\mu \nu }^{a}\right)
+J^{2}\frac{\lambda _{2}}{32}\left( \overline{B}_{\mu \nu
}^{a}-B_{\mu \nu }^{a}\right) ^{2}+\varsigma\,J^4\right)}\,.
\end{equation}
Here, we can appreciate the role of the $\varsigma\, J^4$ term. Upon
integrating over the fields, it becomes clear that we need a
counterterm $\delta\varsigma\, J^4$ to remove the divergent
$J^4$-quantum corrections to $W(J)$. Hence, we need a parameter
$\varsigma$ to absorb this counterterm $\delta\varsigma\,J^4$.
Although it seems that we are introducing a new free parameter into
the action in this manner, $\varsigma$ can be made a unique function
of the coupling constant(s) by requiring a homogenous
renormalization group equation for the effective action, see
\cite{Knecht:2001cc} for applications to the $\lambda\phi^4$ and
Coleman-Weinberg model.

We now define in the usual way
\begin{eqnarray}\label{ea2}
\varphi(x) &=& \frac{\delta  W(J)}{\delta J(x)}\,.
\end{eqnarray}
The original theory (i.e. Yang-Mills) is recovered in the physical
limit $J=0$, in which case we have
\begin{equation}\label{ea3}
    \varphi =
    \frac{i}{4}\Braket{(B_{\mu\nu}^a-\oB_{\mu\nu}^a)F_{\mu\nu}^a}\,.
\end{equation}
If we construct the effective action $\Gamma(\varphi)$, we can thus
study the condensation of the gauge invariant operator
$(B_{\mu\nu}^a-\oB_{\mu\nu}^a)F_{\mu\nu}^a$. The functionals
$\Gamma(\varphi)$ and $W(J)$ are related through a Legendre
transformation
\begin{eqnarray}\label{ea4}
\Gamma(\varphi) &=& W(J) - \int \d^4 x \ J(x)\varphi(x)  \;.
\end{eqnarray}
The vacuum corresponds to the solution of
\begin{eqnarray}
\frac{\p}{\p \varphi}\Gamma(\varphi) &=& 0~(=-J)\;,
\end{eqnarray}
with minimal energy. From now on, we shall restrict ourselves to
space-time independent $\varphi$ and $J$.

In the current situation, we shall have to perform the Legendre
transformation explicitly \cite{Yokojima:1995hy}. Let us give an
illustrative example with a ``toy functional'' $W(J)$
\begin{equation}\label{inv1}
    W(J)=\frac{a_0}{4} J^4+ \frac{g^2}{4}J^4\left(a_1+a_2\ln\frac{J}{\omu}\right)+\textrm{higher order
    terms}\,,
\end{equation}
where $\omu$ is the renormalization scale. Hence
\begin{equation}\label{inv2}
    \varphi=a_0 J^3+ g^2J^3\left(a_1+\frac{a_2}{4}+a_2\ln\frac{J}{\omu}\right)+\textrm{higher order
    terms}\,,
\end{equation}
which leads to
\begin{equation}\label{inv3}
    J=\left(\frac{\varphi}{a_0}\right)^{1/3}\left(1-\frac{g^2}{3a_0}\left(a_1+\frac{a_2}{4}+a_2\ln\frac{(\varphi/a_0)^{1/3}}{\omu}\right)\right)+\textrm{higher order
    terms}\,.
\end{equation}
The trivial vacuum with $\varphi=0$ is of course always recovered,
but there is the possibility for an alternative solution
$\varphi\neq0$, when solving the equation $0=-J=\frac{\p
\Gamma}{\p\varphi}$.

In practice, one can determine $W(J)$ up to the lowest orders in
perturbation theory. $\Gamma(\varphi)$ itself is obtained by
substituting \eqref{inv3} into \eqref{ea4} to reexpress everything
in terms of $\varphi$.

We are now ready to have a look at the effective action in the
\emph{condensed vacuum}. We shall find that the \emph{tree level}
action gets modified in the following way
\begin{eqnarray}\label{inv4}
    \Sigma \to \Sigma'&\equiv& S_{YM}'+\int\d^4x\left[\frac{im}{4}(B-\overline{B})_{\mu\nu}^aF_{\mu\nu}^a
  -\frac{3}{8}%
m^{2}\lambda _{1}\left( \overline{B}_{\mu \nu }^{a}B_{\mu \nu
}^{a}-\overline{G}_{\mu \nu }^{a}G_{\mu \nu }^{a}\right)
+m^{2}\frac{\lambda _{2}}{32}\left( \overline{B}_{\mu \nu
}^{a}-B_{\mu \nu }^{a}\right) ^{2}\right]\nonumber\\
&&+\textrm{higher order terms}\,,
\end{eqnarray}
with
\begin{eqnarray}
m=\left(\frac{\varphi}{a_0}\right)^{1/3}\,,
\end{eqnarray}
since at tree level we only have to take the lowest order term of
\eqref{inv3} with us.

The actual computation of the effective action for the gauge
invariant local composite operator
$(B_{\mu\nu}^a-\oB_{\mu\nu}^a)F_{\mu\nu}^a$ will be the subject of
future work, as this requires a rather large amount of calculations
and the knowledge of yet undetermined renormalization group
functions to two-loop order \cite{Knecht:2001cc,Ford:2009qh}. Anyhow, we
expect that the theory will experience a gauge invariant dimensional
transmutation, leading to
$\Braket{(B_{\mu\nu}^a-\oB_{\mu\nu}^a)F_{\mu\nu}^a}\sim
\Lambda_{QCD}^3$. Further steps towards the effective potential calculation were set in the recent work \cite{Ford:2009qh}.

\sect{The link with perturbative confinement} We did not
substantiate yet the role of the extra parameters $\lambda_{1}$ and
$\lambda_{2}$. We consider the case
\begin{equation}\label{spec}
\lambda_1=\frac{2}{3}\,,\qquad\lambda_2 =0\,.
\end{equation}
Returning for a moment to the Coulomb gauge in the static
case\footnote{Meaning that we formally set ``$\p_0=0$''.}, it is
easy to verify at lowest (quadratic) order that the $(A_0, A_0)$
sector exactly reduces to that of \eqref{1}, by integrating out the
extra fields.

Since we have the freedom to choose the tree level (``classical'')
values for $\lambda_1$ and $\lambda_2$ as we want, we can always
make the confining scenario work by assigning the values
\eqref{spec}. The higher order quantum corrections will consequently
induce perturbative corrections in the couplings $g^2$ and
$\lambda^{abcd}$ to the leading order Cornell potential\footnote{We
shall comment on the role of the tensor coupling $\lambda^{abcd}$
later on in this note.}. At the current time we cannot make more
definite statements about this, as the corresponding renormalization
group functions of $\lambda_1$ and $\lambda_3$ have not yet been
calculated explicitly, see also \cite{Ford:2009qh}. The upshot would of course be
to keep the expansion under control, i.e. to have a reasonably small
expansion parameter. If the dynamically generated mass scale is
sufficiently large, one can readily imagine to have an effective
coupling constant $g^2$ which is relatively small due to asymptotic
freedom. It is perhaps noteworthy to recall the possible emergence
of linear piece of the potential at short distance: restricting to
short distance, i.e. high momentum, might be useful in combination
with asymptotic freedom.

Anyhow, we envisage that the essential nontrivial dynamics would be
buried in the tree level mass parameter (i.e. the nontrivial
condensate $\varphi$), which characterizes an effective action with
confining properties. One can then perform a perturbative weak
coupling expansion around this nontrivial vacuum.

\sect{The static quark potential via the Wilson loop} So far, we
have been looking at the Coulomb gauge to get a taste of the inter
quark potential. However, there is a cleaner (gauge invariant) way
to define the static inter quark potential
$V_{Q\overline{Q}}(\mathbf{r})$. As it is well known,
$V_{Q\overline{Q}}(\mathbf{r})$ can be related to the expectation
value of a Wilson loop, see e.g. \cite{Bali:2000gf,Brown:1979ya}.
More precisely,
\begin{equation}\label{20}
V_{Q\overline{Q}}(\mathbf{r}-\mathbf{r}')=\lim_{T\to\infty}\frac{1}{T}\ln\frac{\mathrm{Tr}\braket{\mathcal{W}}}{\mathrm{Tr}\braket{\mathbf{1}}}\,,
\end{equation}
with the Wilson loop $\cal W$ defined by
\begin{equation}\label{21}
{\cal W}=\mathcal{P} \e^{g\oint_\mathcal{C} A_\mu \d x_\mu}\,,
\end{equation}
where the symbol $\mathcal{P}$ denotes path ordering, needed in the
non-Abelian case to ensure the gauge invariance of
$\mathrm{Tr}\mathcal{W}$. The symbol $\mathbf{1}$ is the unit matrix
corresponding to the representation $R$ of the ``quarks''. Let $t^a$
be the corresponding generators. We shall consider a rectangular
loop $\mathcal{C}$ connecting 2 charges at respective positions
$\mathbf{r}$ and $\mathbf{r}'$, with temporal extension
$T\to\infty$.

To explicitly calculate \eqref{20}, we shall mainly follow
\cite{Fischler:1977yf}. First, we notice that at $T\to\infty$,
$F_{\mu\nu}^2\to0$, i.e. $A_\mu$ becomes equivalent to a pure gauge
potential\footnote{We discard gauge potentials with nontrivial
topology.}, $A_\mu=0$, meaning that we can rewrite the trace of the
Wilson loop as
\begin{equation}\label{22}
\mathrm{Tr}{\cal W}=\mathrm{Tr}\mathcal{P} \e^{g\int
A_0(\mathbf{r},t) \d t-g\int A_0(\mathbf{r}',t) \d t}\,.
\end{equation}
We introduce the current,
\begin{equation}\label{23}
    J_\mu^a(\mathbf{x},t)=
    g\delta_{\mu0}t^a \delta^{(3)}(\mathbf{x}-\mathbf{r})-g\delta_{\mu0}t^a
    \delta^{(3)}(\mathbf{x}-\mathbf{r}')\,,
\end{equation}
to reexpress the expectation value of \eqref{22} as
\begin{equation}\label{24}
    \mathrm{Tr}\braket{{\cal W}}= \frac{\mathcal{P}}{{\cal N}}\int [\d\Phi] \e^{-\Sigma'+\int \d^4x J_\mu^a
    A_\mu^a)}\,,
\end{equation}
with ${\cal N}$ the appropriate normalization factor.

We are now ready to determine the potential explicitly. We limit
ourselves to lowest order, in which case the path ordering is
irrelevant, and we find
\begin{equation}\label{25}
    V_{Q\overline{Q}}(\mathbf{r}-\mathbf{r}')= \frac{1}{\mathrm{Tr}\mathbf{1}}\lim_{T\to\infty}\frac{1}{T}\int
    \frac{\d^4p}{(2\pi)^4}
    \frac{1}{2}J_{\mu}^a(p) D_{\mu\nu}^{ab}(p) J_{\nu}^b(-p)\,,
\end{equation}
with
\begin{equation}\label{26}
    J_\mu^a(p)=2\pi
    g\delta(p_0)(e^{-i\mathbf{p}\cdot\mathbf{r}}-e^{-i\mathbf{p}\cdot\mathbf{r}'})\delta_{\mu0}t^a\,,
\end{equation}
and with
\begin{equation}\label{27}
D_{\mu\nu}^{ab}(p)=D_{\mu\nu}(p)\delta^{ab}\,,\qquad
D_{\mu\nu}(p)=\frac{p^2+m^2}{p^4}\left(\delta_{\mu\nu}-\frac{p_\mu
p_\nu}{p^2}\right)+\frac{\alpha}{p^2}\frac{p_\mu p_\nu}{p^2}\,,
\end{equation}
the gluon propagator. Proceeding with \eqref{25}, we get
\begin{eqnarray}\label{28}
    V_{Q\overline{Q}}(\mathbf{r}-\mathbf{r}')&=& \lim_{T\to\infty}\frac{1}{2T}C_2(R)\int
    \frac{\d^4p}{(2\pi)^4}g^2\delta^2(p_0)
    (2\pi)^2(e^{-i\mathbf{p}\cdot\mathbf{r}}-e^{-i\mathbf{p}\cdot\mathbf{r}'})(e^{i\mathbf{p}\cdot\mathbf{r}}-e^{i\mathbf{p}\cdot\mathbf{r}'})
    D_{00}(p)\nonumber\\
    &=& \lim_{T\to\infty}\frac{1}{2T}C_2(R) g^22\pi\delta(0)\int
    \frac{\d^3\mathbf{p}}{(2\pi)^3}(e^{-i\mathbf{p}\cdot\mathbf{r}}-e^{-i\mathbf{p}\cdot\mathbf{r}'})(e^{i\mathbf{p}\cdot\mathbf{r}}-e^{i\mathbf{p}\cdot\mathbf{r}'})
    D_{00}(p)_{p_0=0}\nonumber\\
    &=&-g^2C_2(R)\int
    \frac{\d^3\mathbf{p}}{(2\pi)^3}\frac{\mathbf{p}^2+m^2}{\mathbf{p}^4}-g^2C_2(R)\int
    \frac{\d^3\mathbf{p}}{(2\pi)^3}\frac{\mathbf{p}^2+m^2}{\mathbf{p}^4}\e^{i\mathbf{p}(\mathbf{r}-\mathbf{r}')}\,.
\end{eqnarray}
We used that $\displaystyle \lim_{T\to\infty} T=\displaystyle
 \lim_{T\to\infty}\int_{-T/2}^{T/2}\d t=2\pi\delta(0)$. The first
term of \eqref{28} corresponds to the (infinite) self energy of the
external charges \cite{Fischler:1977yf}, so we can neglect this term
to identify the interaction energy, which yields after performing
the Fourier integration
\begin{eqnarray}\label{29}
    V_{Q\overline{Q}}(\mathbf{r}-\mathbf{r}')&=&\frac{g^2
    C_2(R)}{8\pi}m^2\vert \mathbf{r}-\mathbf{r}'\vert-\frac{g^2C_2(R)}{4\pi}\frac{1}{\vert
    \mathbf{r}-\mathbf{r}'\vert}\,.
\end{eqnarray}
We nicely obtain a Cornell potential, with the string tension in
representation $R$ given by $\sigma(R)= \frac{g^2}{8\pi}C_2(R)m^2$.
Notice that the so-called Casimir scaling \cite{Bali:2000un} of
$\sigma(R)$ is straightforwardly fulfilled, at least at the
considered order.

If we
consider our model in a specific gauge, for example the Landau gauge, we
see the presence of a $\frac{1}{p^4}$ singularity in the (tree
level) gluon propagator \eqref{27}. Actually, it was already argued in
\cite{West:1982bt} that such pole would induce the area law of the
Wilson loop, if present in \emph{some} gauge. In the Landau gauge in
particular, lattice data have already ruled out since long such a
highly singular gluon propagator, see \cite{Cucchieri:2007rg} for a
recent numerical analysis.

A first observation is that we presented only a lowest order calculation, based on the tree level gluon propagator. We did not consider quantum corrections, on neither the Wilson loop's expectation value nor gluon propagator. A more sophisticated treatment would also have to take into account that our naive string tension $\sigma$, related to the condensate $\braket{B-\overline{B}}F$, will run with the scale. This would ask for a renormalization group improved treatment. We already mentioned in the introduction that the string tension at short distance (large energy scale) does not have to
concur with the one at large distances (small energy scale) \cite{Andreev:2006ct,Zakharov:2007zzb}.

We must also remind that most gauges, in particular, the Landau gauge, are plagued by the Gribov copy problem, which also influence the infrared dynamics of a gauge theory \cite{Gribov:1977wm,Dudal:2008sp}. The latter problem can be overcome as we are not obliged to work in the Landau gauge, since we
have set up a gauge invariant framework. In most other gauges, it is
not even known how to tackle e.g. the gauge copy problem in a more
or less tractable way, or there are no copies at all in certain
gauges\footnote{Some of these gauges then suffer from other
problems.}. As an example of the latter gauges, let us impose the
planar gauge \cite{Leibbrandt:1987qv} via a gauge fixing term
$S_{gf}~=~\int \d^4x \frac{1}{2n^2} n\cdot A \p^2 n\cdot A$. The
gluon propagator becomes a bit complicated
\begin{eqnarray}
D_{\mu\nu}^{ab}(p)&=&\delta^{ab}\left(\frac{p^2+m^2}{p^4}\delta_{\mu\nu}+m^2\frac{p^2+m^2}{p^4}\frac{n^2}{(p\cdot
n)^2}\frac{p_\mu p_\nu}{p^2}-\frac{p^2+m^2}{p^4}\frac{n_\mu
p_\nu}{p\cdot n}-\frac{(p^2+m^2)^2}{p^6}\frac{p_\mu n_\nu}{p\cdot
n}\right)\,,
\end{eqnarray} nevertheless the result
\eqref{29} is recovered, after some algebra.

\sect{Symmetry breaking pattern}
We already mentioned the useful
supersymmetry $\delta^{(2)}$, which is however broken if
$\Braket{(B_{\mu\nu}^a-\oB_{\mu\nu}^a)F_{\mu\nu}^a}\neq0$ (i.e.
\mbox{$m\neq0$}). Hence, we should worry about the emergence of an
extra (undesired) massless degree of freedom: the associated
Goldstone fermion\footnote{Not boson, as $\delta_2$ transforms
bosons into fermions and vice versa.}. The situation is however more
complicated than this. The starting action $S_{YM}'$ enjoys the
following set of (nilpotent) supersymmetries
\begin{eqnarray}\label{gold1}
    \delta^{(1)}&=&\int \d^4x \left(B_{\mu\nu}^a\frac{\delta}{\delta G_{\mu\nu}^a}-\oG_{\mu\nu}^a\frac{\delta}{\delta \oB_{\mu\nu}^a}\right)\,,\qquad\delta^{(3)}~=~\int \d^4x \left(\oB_{\mu\nu}^a\frac{\delta}{\delta G_{\mu\nu}^a}-\oG_{\mu\nu}^a\frac{\delta}{\delta
    B_{\mu\nu}^a}\right)\,,\nonumber\\
\delta^{(2)}&=&\int \d^4x \left(\oB_{\mu\nu}^a\frac{\delta}{\delta
\oG_{\mu\nu}^a}+G_{\mu\nu}^a\frac{\delta}{\delta
B_{\mu\nu}^a}\right)\,,\qquad\delta^{(4)}~=~\int \d^4x
\left(B_{\mu\nu}^a\frac{\delta}{\delta
\oG_{\mu\nu}^a}+G_{\mu\nu}^a\frac{\delta}{\delta
    \oB_{\mu\nu}^a}\right)\,,
\end{eqnarray}
in addition to the bosonic symmetries generated by
\begin{eqnarray}\label{gold2}
    \Delta^{(1)}&=&\int \d^4x \left(B_{\mu\nu}^a\frac{\delta}{\delta B_{\mu\nu}^a}-\oB_{\mu\nu}^a\frac{\delta}{\delta \oB_{\mu\nu}^a}\right)\,,\qquad\Delta^{(2)}~=~\int \d^4x \left(G_{\mu\nu}^a\frac{\delta}{\delta G_{\mu\nu}^a}-\oG_{\mu\nu}^a\frac{\delta}{\delta
    \oG_{\mu\nu}^a}\right)\,.
\end{eqnarray}
It appears that a nonvanishing
$\Braket{(B_{\mu\nu}^a-\oB_{\mu\nu}^a)F_{\mu\nu}^a}$ results in the
dynamical breakdown of the continuous symmetries
$\delta^{(1),(2),(3),(4)}$ and $\Delta^{(1)}$. Though, a little more
care is needed. Not all the breakings are independent, as one checks
that
\begin{eqnarray}\label{gold3}
\delta^{(1)-(3)}~\equiv~ \delta^{(1)}-\delta^{(3)}
\,,\qquad\delta^{(2)-(4)}~\equiv~
\delta^{(2)}-\delta^{(4)}\,,\qquad\Delta^{(1)}\,,
\end{eqnarray}
are clearly dynamically broken for $\braket{(B-\oB)F}\neq0$, since
can write
\begin{eqnarray}\label{gold3bis}
\Braket{(B_{\mu\nu}^a-\oB_{\mu\nu}^a)F_{\mu\nu}^a}=\Braket{\delta^{(1)-(3)}\left[G_{\mu\nu}^aF_{\mu\nu}^a\right]}=-\Braket{\delta^{(2)-(4)}\left[\oG_{\mu\nu}^aF_{\mu\nu}^a\right]}=\Braket{\Delta^{(1)}\left[(B_{\mu\nu}^a+\oB_{\mu\nu}^a)F_{\mu\nu}^a\right]}\,,
\end{eqnarray}
while
\begin{eqnarray}\label{gold4}
\delta^{(1)+(3)}~\equiv~ \delta^{(1)}+\delta^{(3)}
\,,\qquad\delta^{(2)+(4)}~\equiv~
\delta^{(2)}+\delta^{(4)}\,,\qquad\Delta^{(2)}\,,
\end{eqnarray}
are still conserved.

If a nonzero value of
$\Braket{(B_{\mu\nu}^a-\oB_{\mu\nu}^a)F_{\mu\nu}^a}$ is dynamically
favoured, 2 Goldstone fermions and 1 Goldstone boson seem to enter
the physical spectrum. As this would be a serious
problem\footnote{These extra particles carry no color, so there is
no reason to expect that these would be confined or so, thereby
removing themselves from the physical spectrum.}, we need to find a
way to remove these from the spectrum. A typical way to kill
unwanted degrees of freedom is by imposing constraints on the
allowed excitations. Consistency is assured when this is done by
using symmetry generators to restrict the physical subspace. First,
we have to identify the suitable operators to create/annihilate the
Goldstone particles. As it is well known, these are provided by the
Noether currents corresponding to \eqref{gold3}, which can be
derived from the action $S_{YM}'$. We obtain
\begin{eqnarray}\label{gold5}
j^{(1)-(3)}_\mu&=&-B_{\alpha\beta}^a
D_{\mu}^{ab}\oG_{\alpha\beta}^b+\oG_{\alpha\beta}^a
D_{\mu}^{ab}B_{\alpha\beta}^b+\oB_{\alpha\beta}^a
D_{\mu}^{ab}\oG_{\alpha\beta}^b-\oG_{\alpha\beta}^a
D_{\mu}^{ab}\oB_{\alpha\beta}^b\;,\nonumber\\
j^{(2)-(4)}_\mu&=&\oB_{\alpha\beta}^a
D_{\mu}^{ab}G_{\alpha\beta}^b-G_{\alpha\beta}^a
D_{\mu}^{ab}\oB_{\alpha\beta}^b-B_{\alpha\beta}^a
D_{\mu}^{ab}G_{\alpha\beta}^b+G_{\alpha\beta}^a
D_{\mu}^{ab}B_{\alpha\beta}^b\,,
\end{eqnarray}
after a little algebra. Let us now define what physical operators
are. First of all, they are expected to be gauge
invariant\footnote{Or more precisely, BRST closed but not exact,
after fixing the gauge.}. Secondly, based on $\Delta^{(2)}$ we can
also introduce a $\mathcal{G}$-ghost charge, with
$\mathcal{G}(G_{\mu\nu}^a)=+1$, $\mathcal{G}(\oG_{\mu\nu}^a)=-1$,
and demand that physical operators are $\mathcal{G}$-neutral. In
addition, we also can request invariance w.r.t. $\delta^{(1)+(3)}$
and $\delta^{(2)+(4)}$.

Let us mention the following useful relations
\begin{eqnarray}\label{gold6}
    \delta^{(1)+(3)}j_\mu^{(2)-(4)}&=&\delta^{(2)+(4)}j_\mu^{(1)-(3)}~=~ 2(\oB_{\alpha\beta}^a
D_{\mu}^{ab}B_{\alpha\beta}^b-B_{\alpha\beta}^a
D_{\mu}^{ab}\oB_{\alpha\beta}^b)\neq0\;,\nonumber\\
\delta^{(1)+(3)}j^{(1)-(3)}&=&\delta^{(2)+(4)}j^{(2)-(4)}~=~0\,.
\end{eqnarray}
The currents $j_\mu^{(2)-(4)}$ or $j_\mu^{(1)-(3)}$ are thus not
physical operators. Although gauge invariant, \eqref{gold6} tells us
these are not $\delta^{(1)+(3)}$ or $\delta^{(2)+(4)}$ invariant.
Moreover, since $\mathcal{G}(j_\mu^{(2)-(4)})=+1$, and
$\mathcal{G}(j_\mu^{(1)-(3)})=-1$, also the $\mathcal{G}$-neutrality
is not met.

We can assure $\mathcal{G}$-neutrality by e.g. taking a product
$j^{(2)-(4)}j^{(1)-(3)}$, but this does not ensure
$\delta^{(1)+(3)}$ or $\delta^{(2)+(4)}$ invariance, which can be
easily checked using \eqref{gold6}.

Concerning the current $k_\mu$ associated with $\Delta^{(1)}$, we
find
\begin{eqnarray}\label{gold7}
k_\mu&=&-B_{\alpha\beta}^a
D_\mu^{ab}\oB_{\alpha\beta}^a+\oB_{\alpha\beta}^a
D_\mu^{ab}B_{\alpha\beta}^a\,,
\end{eqnarray}
hence
\begin{eqnarray}\label{gold8}
\delta^{(1)+(3)}k_\mu&=&B_{\alpha\beta}^a
D_\mu^{ab}\oG_{\alpha\beta}^a-\oG_{\alpha\beta}^a
D_\mu^{ab}B_{\alpha\beta}^a+\oG_{\alpha\beta}^a
D_\mu^{ab}\oB_{\alpha\beta}^a-\oB_{\alpha\beta}^a
D_\mu^{ab}\oG_{\alpha\beta}^a~\neq~0\;,\nonumber\\
\delta^{(2)+(4)}k_\mu&=&-G_{\alpha\beta}^a
D_\mu^{ab}\oB_{\alpha\beta}^a+\oB_{\alpha\beta}^a
D_\mu^{ab}G_{\alpha\beta}^a-B_{\alpha\beta}^a
D_\mu^{ab}G_{\alpha\beta}^a+G_{\alpha\beta}^a
D_\mu^{ab}B_{\alpha\beta}^a~\neq~0\,.
\end{eqnarray}
Since the symmetries we are using are not unrelated, it is evidently
no surprise that $k_\mu$, $j_\mu^{(2)-(4)}$ and $j_\mu^{(1)-(3)}$
are transformed into each other. The question remains however
whether we can build combinations\footnote{These combinations may of
course contain other operators too.} of these which enjoy all the
necessary invariances? Let us try to construct one, starting from
$j^{(2)-(4)}$. We shall use a more symbolic notation. It can be
checked that e.g.
\begin{equation}\label{gold9}
    \delta^{(2)+(4)}\left(\oG
    j^{(2)-(4)}+(B+\oB)K-Gj^{(1)-(3)}\right)~=~0\,,
\end{equation}
but
\begin{equation}\label{gold10}
    \delta^{(1)+(3)}\left(\oG j^{(2)-(4)}+(B+\oB)K-Gj^{(1)-(3)}\right)=-4\oG
    k-2(B+\oB)j^{(1)-(3)}\,.
\end{equation}
So far, we have been unable to construct suitable invariant operators. We are lead to believe that this is generally true, in return we could
state that the Goldstone modes can be expelled from the spectrum. An explicit proof is however lacking hitherto.

\sect{A few words on the tensor coupling $\lambda^{abcd}$} In the
massless case, the precise value of the tensor coupling
$\lambda^{abcd}$ is irrelevant, as it cannot influence the dynamics
of the (physical) Yang-Mills  sector of the theory as explained
above. However, when studying the effective action for
$\varphi=\braket{(B-\oB)F}$, $\lambda^{abcd}$ plays a role. We might
see this as a drawback, as then a new independent coupling would
enter the game. As our setup was to deal with confinement in usual
gauge theories with a single gauge coupling $g^2$, we would like to
retain solely $g^2$ as the relevant parameter. This can be nicely
accommodated for by invoking the renormalization group equations to
reduce the number of couplings. In the presence of multiple
couplings, one can always opt to choose a primary coupling and
express the others in term of this one. For consistency, no
sacrifices should be made w.r.t. the renormalization group
equations, therefore we shall search for a fix point
$\lambda_*^{abcd}(g^2)$, such that
$\mu\frac{\p}{\p\mu}\lambda_*^{abcd}=0$.

We recall the result of \cite{Capri:2006ne}, where it was
calculated, using dimensional regularization $(d=4-2\epsilon)$ and
using the $\MSbar$ scheme, that
\begin{eqnarray}\label{betaabcdbis}
    \mu\frac{\p}{\p\mu}\lambda^{abcd}&=&-2\varepsilon\lambda^{abcd}+\left[ \frac{1}{4} \left(
\lambda^{abpq} \lambda^{cpdq} + \lambda^{apbq} \lambda^{cdpq} +
\lambda^{apcq} \lambda^{bpdq}
+ \lambda^{apdq} \lambda^{bpcq} \right) \right. \nonumber \\
&& \left. -~ 12 C_A \lambda^{abcd} a ~+~ 8 C_A f^{abp} f^{cdp} a^2
~+~ 16 C_A f^{adp} f^{bcp} a^2 ~+~ 96 d_A^{abcd} a^2
\right]+\ldots\,,
\end{eqnarray}
with $a=\frac{g^2}{16\pi^2}$, and we also rescaled
$\lambda^{abcd}\to\frac{1}{16\pi^2}\lambda^{abcd}$. We clearly
notice that $\lambda^{abcd}=0$ is not a fixed point of this
renormalization group equation.  We must thus look out for an
alternative fixed point $\lambda_*^{abcd}\neq0$.

We shall restrict ourselves to the simplest case: we take $SU(2)$ as
gauge group, and only consider gauge fields in the adjoint
representation. Doing so, we can simplify \eqref{betaabcdbis} a bit
by explicitly computing the completely symmetric rank 4 tensor
$d_A^{abcd}$ \cite{vanRitbergen:1998pn}, and by looking for tensor
structures that can be used to construct a rank 4 tensor consistent
with the constraints \eqref{jacobigen} and \eqref{abcd}.

The generators of the adjoint representation of $SU(2)$, are given
by    $(t^a)_{bc}=i\varepsilon^{abc}$. We can compute $d_A^{abcd}$,
which is defined by means of a symmetrized trace $\mbox{STr}$ as
\begin{eqnarray} \label{gen3}
\!\!\!\!\!\!  d^{abcd}_A &=&\mbox{STr}\left(t^at^bt^ct^d\right)
~=~\left[\delta^{ab}\delta^{cd}+\delta^{ad}\delta^{bc}\right]_{\mathrm{symmetrized\,w.r.t.}\,\{a,b,c,d\}
}
  ~=~\frac{2}{3}\left(\delta^{ab}\delta^{cd}+\delta^{ac}\delta^{bd}+\delta^{ad}\delta^{bc}\right)\,.
\end{eqnarray}
Moreover, we can also simplify the other tensor appearing in
\eqref{betaabcdbis}, namely ($C_A=2$)
\begin{eqnarray}\label{gen4}
 8 C_A f^{abp} f^{cdp} a^2+16 C_A f^{adp} f^{bcp}
 a^2&=&-16\delta^{ac}\delta^{bd}-16\delta^{ad}\delta^{bc}+32\delta^{ab}\delta^{cd}\,.
\end{eqnarray}
Using the constraints \eqref{jacobigen} as definition of any
building block of our tensor $\lambda_\ast^{abcd}$, one can check
that the following rank 4 color tensors are suitable (linearly
independent) candidates
\begin{eqnarray}
\label{gen5a} {\cal O}_1^{abcd}&=&\delta^{ab}\delta^{cd}\,,\qquad
{\cal O}_2^{abcd}~=~\delta^{ac}\delta^{bd}+\delta^{ad}\delta^{bc}\,.
\end{eqnarray}
Clearly, $d_A^{abcd}$ and the tensor \eqref{gen4} are particular
linear combinations of the tensors in \eqref{gen5a}. We now propose
\begin{equation}\label{gen11}
    \lambda_f^{abcd}(a)=y_1{\cal
O}_1^{abcd}a+y_2{\cal O}_2^{abcd}a\qquad\qquad y_i\in\mathbb{R}\,,
\end{equation}
and we demand that the l.h.s. of \eqref{betaabcdbis} vanishes when
\eqref{gen11} is substituted into it, with $\epsilon=0$. This leads
to
\begin{eqnarray}\label{gen13}
    \left\{\begin{array}{ccc}
    y_1&\approx&67.6\\
y_2&\approx&-43.6
           \end{array}
    \right.\,,\qquad\qquad\left\{\begin{array}{ccc}
    y_1&\approx&28.4\\
y_2&\approx&-4.4
           \end{array}
    \right.\,.
\end{eqnarray}
We conclude that the renormalization group equation
$\mu\frac{\p}{\p\mu}\lambda^{abcd}=\beta^{abcd}=0$ possesses a fixed
point in $d=4$, at least at 1-loop for the gauge group $SU(2)$ in
the presence of only gauge fields.

We end this note by briefly returning to the issue of
$\frac{1}{q^2}$ power corrections. In
\cite{Gubarev:2000eu,Gubarev:2000nz}, these were related to (part
of) the dimension two condensate $\braket{A^2_{\min}}=
(VT)^{-1}\braket{\min_{g\in SU(N)}\int \d^4x (A_\mu^g)^2}$. The
nonlocal operator $A^2_{\min}$ reduces to $A^2$ in the Landau gauge,
hence the interest in this gauge
\cite{Gubarev:2000eu,Gubarev:2000nz}. Although the mechanism
discussed in this Letter might seem to be completely different, this
is however not the case. The nonperturbative mass scale, set by the
condensation of the gauge invariant operator \eqref{ea3}, will also
fuel a nonvanishing $A^2$ condensate in the Landau gauge, i.e.
$\braket{A^2}\propto m^2$, already in a perturbative loop expansion.
As such, at least part of the nonperturbative information stored in
$\braket{A^2}$ could be attributed to the gauge invariant condensate
introduced in this work.

\section*{Acknowledgments}
D.~Dudal is grateful to R.~Jackiw for useful discussions. D.~Dudal
is supported by the Research-Foundation Flanders. This work is
supported in part by funds provided by the US Department of Energy
(DOE) under cooperative research agreement DEFG02-05ER41360.

\end{document}